\documentclass{aa}
\usepackage{graphicx}
\usepackage{txfonts}
\usepackage{psfrag}\usepackage{color}
\usepackage{natbib}
\bibpunct{(}{)}{;}{a}{}{,} 
\usepackage{bm}

\begin{document}
\title{Interplay between chemistry and dynamics in embedded protostellar disks}
\titlerunning{}

\author{C.~Brinch\inst{1,2} \and J.~K.~J{\o}rgensen\inst{1,2}}
\institute{
Niels Bohr Institute, University of Copenhagen,
Juliane Maries Vej 30, DK-2100 Copenhagen \O, Denmark
\and
Centre for Star and Planet Formation and Natural History Museum of Denmark, University of
Copenhagen, \O ster Voldgade 5–7, DK-1350 Copenhagen K., Denmark\\
\email{brinch@nbi.ku.dk}
}
\date{Received  / Accepted }
\abstract{A fundamental part of the study of star formation is to place young stellar objects in an evolutionary sequence. Establishing a robust evolutionary classification scheme allows us not only to understand how the Sun was born but also to predict what kind of main sequence star a given protostar will become. Traditionally, low-mass young stellar objects are classified according to the shape of their spectral energy distributions. Such methods are however prone to misclassification due to degeneracy and do not constrain the temporal evolution. More recently, young stellar objects have been classified based on envelope, disk, and stellar masses determined from resolved images of their continuum and line emission at submillimeter wavelengths.}
{Through detailed modeling of two Class I sources, we aim at determining accurate velocity profiles and explore the role of freeze-out chemistry in such objects.}
{We present new Submillimeter Array observations of the continuum and HCO$^+$ line emission at 1.1 mm toward two protostars, IRS~63 and IRS~43 in the Ophiuchus star forming region. The sources are modeled in detail using dust radiation transfer to fit the SED and continuum images and line radiation transfer to produce synthetic position-velocity diagrams. We use a $\chi^2$ search algorithm to find the best model fit to the data and to estimate the errors in the model variables.}
{Our best fit models present disk, envelope and stellar masses as well as the HCO$^+$ abundance and inclination of both sources. We also identify a ring structure with a radius of about 200 AU in IRS~63.}
{We find that freeze-out chemistry is important in IRS~63 but not for IRS~43. We show that the velocity field in IRS~43 is consistent with Keplerian rotation. Due to molecular depletion, it is not possible to draw a similar conclusion for IRS~63. We identify a ring shaped structure in IRS~63 on the same spatial scale as the disk outer radius. No such structure is seen in IRS~43.}

\maketitle

\section{Introduction}
Gas dynamics is an all important diagnostic tool in the field of star formation. It is governed by the gravitational pull from the newly formed star as well as the conservation of the angular momentum carried by the gas itself. Spectroscopic observations of the gas surrounding young stars tell us about the current accretion rate, the amount of turbulence, the presence and importance of outflows, and even the evolutionary stage of young stellar objects: in the earliest protostellar stages gas will predominantly move along radial trajectories as the cloud collapses but will eventually settle down in Keplerian orbits in the newly formed disk. To derive meaningful information from  spectroscopic observations it is important to understand the distribution of molecules, that is, the chemistry. Chemical depletion, whether due to freeze-out, dissociation, or gas-phase reactions, can mask out important dynamical regions, e.g., the disk, and thus hide the signatures of accretion or rotation. On the other hand, if the depletion can be accurately determined from detailed modeling of the gas dynamics, we can get valuable information on the chemistry as a function of time~\citep[e.g., review by][]{2007prpl.conf..751B}. 

Protostellar evolution is often described by a series of distinct evolutionary pictures associated with the classification scheme of \citet{Lada1984,1987IAUS..115....1L} and \citet{Andre1993}. In reality, the evolution is a smooth, continuous process where the protostellar envelope is being accreted while the star and the protoplanetary disk grows. It is mostly agreed upon that the classes 0, I, II, and III represent a monotonous time progression~\citep[e.g.,][]{EvansII:2008wg}. There is, however, currently no consensus about how sources of a single class can be sorted with respect to their absolute age. Attempts have been made \citep[e.g.,][]{1993ApJ...413L..47M} to introduce a continuous parametrization of protostellar evolution, usually based on properties of the spectral energy distribution (SED). 

\citet{Robitaille2006} recognized that all the various classifications that are based on the spectral index do not necessarily distinguish between objects that are physically and hence evolutionary distinct. For example, the same source can be classified differently depending on the viewing angle. They introduced an analogous classification scheme, where classes are denoted ``Stage'', that is based on physical properties, such as mass and accretion rate. The main difficulty of classifying stars based on physical properties is that a modeling effort is always, to some extent, required, which is both time consuming and labor intensive as opposed to classifications that are based on apparent properties (such as the shape of the SED) which are easily automated. 

\citet{Jorgensen:2009bx} observed a sample of Class 0 and I sources and looked for evolutionary tracers based on observable properties such as envelope mass derived from the single dish (low resolution) submillimeter flux, disk mass derived from the high resolution compact flux, and stellar luminosity derived from the SED. One conclusion of this study was that the disk mass does not change with decreasing envelope mass, that is, it is constant throughout the Class 0 and Class I stages. The question is however, how robust these derived values are compared to the ones obtained through detailed radiative transfer modeling of the data. Such modeling is non-trivial, however. A model should provide a description of the H$_2$ density, the gas- and dust temperature, the molecular abundance, and the velocity field that need to be constrained by multi-wavelength line and continuum observations. Thanks to space missions like the Spitzer Space Telescope, ISO, and Herschel Space Observatory, as well as ground-based sub-millimeter continuum surveys large databases of young stars are becoming available. In addition, near-future observations with the Atacama Large Millimeter/submillimeter Array (ALMA) will provide a high degree of detail on individual sources.

In this paper, we present an extended analysis of two Class~I protostellar sources in the Ophiuchus star-forming region\footnote{In this paper we adopt a distance of 125 pc to Ophiuchus \citep{1989A&A...216...44D} to be consistent with previous work.}, IRS~43 and IRS~63, from the sample of \citet{Jorgensen:2009bx}. These two sources were chosen because they appear to be at a similar evolutionary stage. Nevertheless, they also show peculiar differences, namely that IRS~63 has weak HCO$^+$ lines on top of a strong continuum while IRS~43 has strong lines on top of a weak continuum. We present additional high-angular resolution spectral line observations in the sub-millimeter and interpret these in the context of a detailed continuum and line radiative transfer model to asses their dynamical and chemical structure. Finally we discuss the exactness of simpler, more generic observations, with a particular eye on the perspectives opened-up by near-future ALMA observations.

\section{Data}\label{data}
Observations of IRS~43 (IRAS 16244-2434) and IRS~63 (IRAS 16285-2355) were made using the Submillimeter Array~\citep[SMA,][]{2004ApJ...616L...1H}\footnote{The Submillimeter Array is a joint project between the Smithsonian Astrophysical Observatory and the Academia Sinica Institute of Astronomy and Astrophysics, and is funded by the Smithsonian Institution and the Academia Sinica.} in extended configuration in July  2008 as a follow-up project to the PROSAC campaign~\citep{Jorgensen:2007gt,Jorgensen:2009bx}. Both sources were observed during a single track using all 8 telescopes, providing projected baselines between 16 k$\lambda$ and 171 k$\lambda$. We used the same spectral set-up as was used in the original compact configuration track, namely one chunk of 512 channels centered on the HCO$^+$ $J=$ 3--2 line (267.557648 GHz) providing a spectral resolution of 0.23 km s$^{-1}$. Another high resolution chunk was placed over the HCN $J=$3--2 line, but this line is undetected for both sources in the extended configuration track. The remaining bandwidth was used to measure the continuum at 1.1 millimeter. All calibration and data reduction were done in IDL using the MIR package~\citep{qi2005} and the data were combined with the compact configuration data using MIRIAD~\citep{Sault1995}. The combined datasets have synthesized beams of 1.2$'' \times$ 1.1$''$ and 1.7$'' \times$ 1.4$''$ for IRS~63 and IRS~43, respectively. For IRS~63, the RMS noise is  2.9 mJy~beam$^{-1}$ for the continuum and 0.2 Jy~beam$^{-1}$~channel$^{-1}$ for the HCO$^+$ line. For IRS~43, the RMS for the continuum and the HCO$^+$ line is 1.7 mJy~beam$^{-1}$ and 0.2 Jy~beam$^{-1}$~channel$^{-1}$.

In this paper we also make use of archival data of both sources. These data include images from the SCUBA legacy catalogues~\citep{DiFrancesco:2008di}, the 2MASS survey~\citep{2006AJ....131.1163S}, a Spitzer/IRS spectrum of IRS~63 from the c2d legacy project~\citep{Evans:2003wg}, $J$, $H$, and $K$ band photometry of IRS~43~\citep{Haisch:2002wf}, as well as 1.1 mm fluxes from CSO/Bolocam survey~\citep{Young:2006vg}. In addition, we use a 55--200 $\mu$m PACS spectrum of IRS~63 from the Herschel Space Observatory DIGIT key program \citep{2013ApJ...770..123G}.

\section{Results} \label{results}
\begin{figure}
   \centering
      \includegraphics[width=4.4cm]{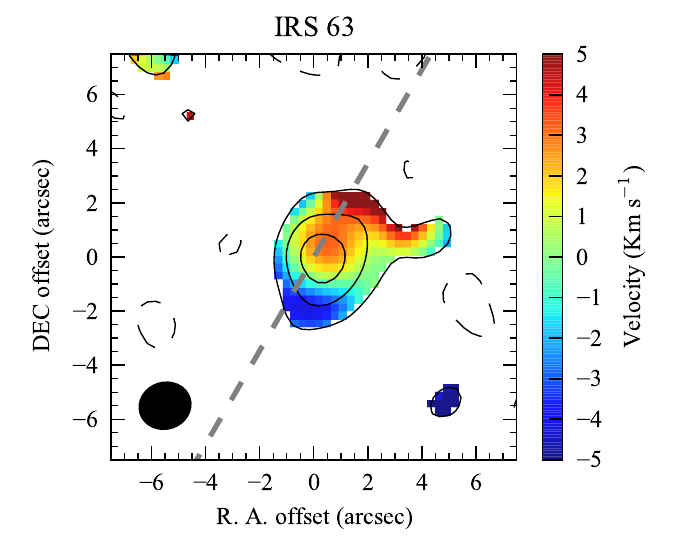}
      \includegraphics[width=4.4cm]{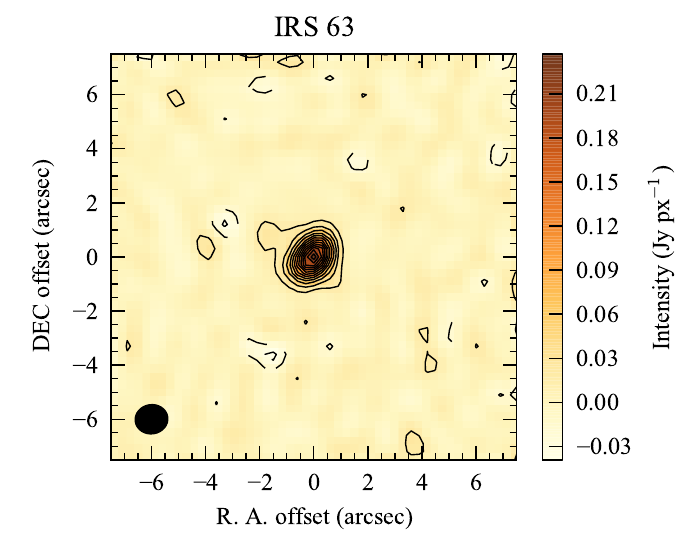}  \\
      \includegraphics[width=4.4cm]{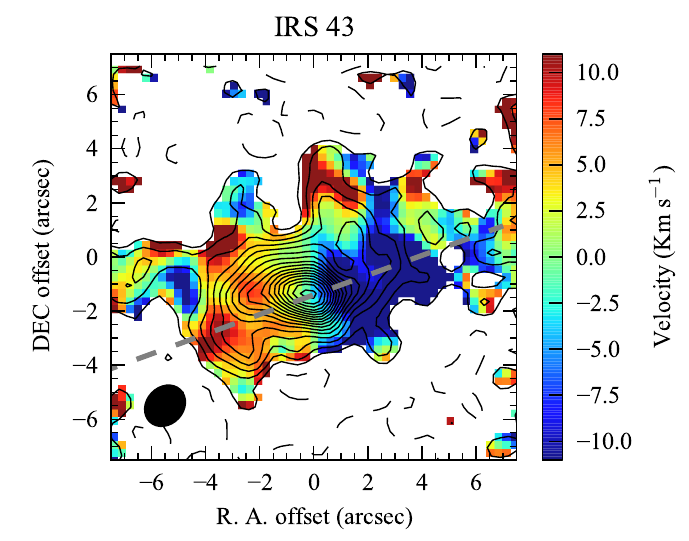}
      \includegraphics[width=4.4cm]{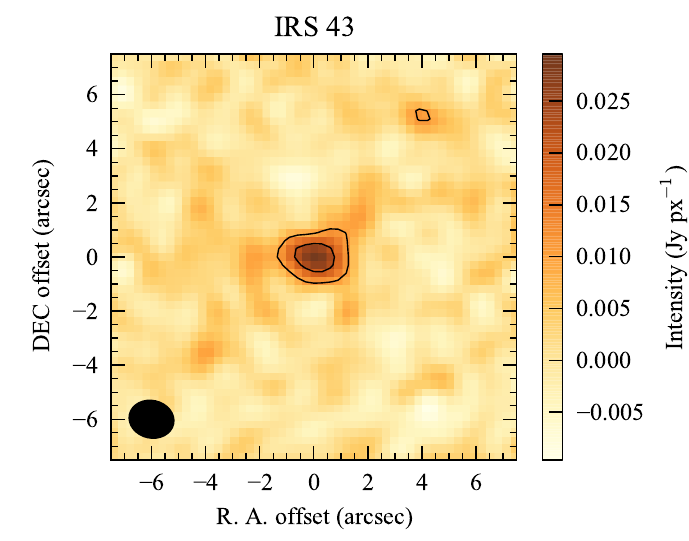}  \\
   \caption{Moment 0 (contour) and 1 (color) maps of the HCO$^+$ J=3-2 line (left) and 1.1 mm continuum image (right) of IRS~63 and IRS~43. The beam is shown by the black ellipses. The dashed grey line marks the direction of the velocity gradient.}
\label{dataplot}
\end{figure}

Figure~\ref{dataplot} shows reconstructed images of the HCO$^+$ line and the continuum emission as observed by the SMA.  Natural weighting of the visibilities was used for the reconstruction in both cases. While the continuum is seen to be only marginally resolved, the HCO$^+$ is well resolved for both sources. The line is considerably brighter in IRS~43 than in IRS~63 and vice versa for the continuum. Both sources show a clear velocity gradient. 

\begin{figure}
   \centering
      \includegraphics[width=8.5cm]{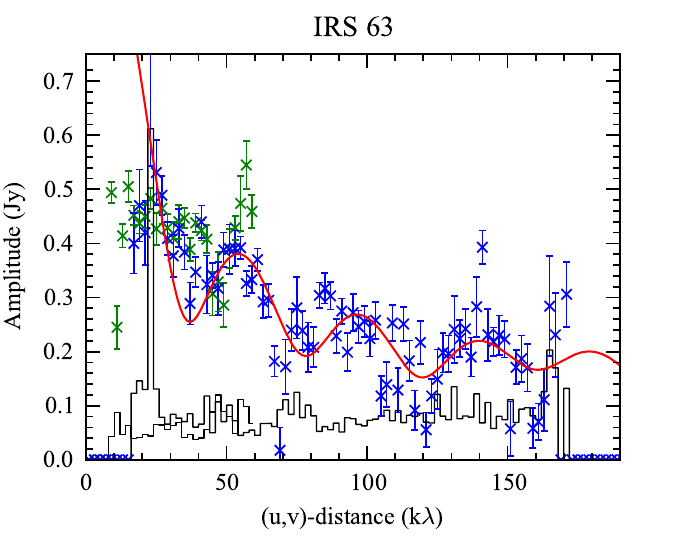}
   \caption{Continuum visibility amplitudes of IRS~63 at 1.1 mm. The green markers are the visibilites covered by the PROSAC compact configuration track while the blue markers show the data covered by the extended configuration track. The black histogram shows the zero-signal expectation. The red curve shows the Fourier transform of a thin ring with a radius of 180 AU.}
\label{ring}
\end{figure}

The structure of IRS~63 was studied by \citet{2008A&A...481..141L} who used the PROSAC compact configuration data from the SMA to derive disk and envelope masses. The compact configuration data have projected baselines up to 60 k$\lambda$ corresponding to a linear scale of about 440 AU. The disk is unresolved and therefore only the disk mass is constrained on these scales. In our extended configuration data with baselines of 171 k$\lambda$ (linear scales of 150 AU) the disk is resolved, however. Figure~\ref{ring} shows the visibility amplitudes of the continuum at 1.1 mm, with the symbols in green showing the visibilities covered by the compact configuration data and symbols in blue showing the visibilities covered by the extended configuration data. The extended configuration data suggest the presence of a face-on ring structure with an approximate diameter of 2.9$''$ equivalent to a linear radius of 180 AU at the distance of IRS~63. Figure~\ref{ring} also shows the Fourier transform of such a ring (with an arbitrary flux) plotted in red. The data obviously also contain the signal from the other components of circumstellar material and therefore the vertical offset of the red curve is not meant as a fit to the data but only as a guide to the eye. The fact that the nulls appear sharper in the data than in our Fourier transformed ring is due the fact that we use an infinitely thin (1D) ring whereas in reality the ring is probably a bit extended and with fuzzy edges. The extended configuration continuum amplitudes of IRS~43, which has a much weaker continuum signature than IRS~63, do not show a similar ring pattern, implying either that either no such structure exists in IRS~43 or, more likely, that the disk in IRS~43 is oriented in such a way that this structure, whatever it may be, is not seen as a ring.

\section{Modeling}\label{models}
\subsection{The continuum}
In order to proceed, physical models of our two sources are needed. Although a physical model of IRS63 was presented by \citet{2008A&A...481..141L}, we have chosen to redo the modeling for two reasons. First of all, we have much more data to constrain the model. Secondly, the data were fit-by-eye by \citet{2008A&A...481..141L}, whereas we are able to do a systematic search of the parameter space and provide error bars on our best fit parameter values. As mentioned above, we use a simple three component model to describe the two YSOs, a protostar, a disk, and an envelope. The star is described by its surface temperature and the stellar radius, which together give its total luminosity. We use the standard description of a 2D disk in hydrostatic equilibrium parameterized by
\begin{eqnarray}\label{eq:disk}
\rho_{disk}(R,z) = \frac{1}{\sqrt{2\pi}H}\Sigma (R) \cdot e^{-\left (\frac{z}{2H}\right )^2}
\end{eqnarray}
where,
\begin{eqnarray}
&\Sigma(R)& = \Sigma_0 (R/R_0)^{-1}\\
&H(R)& = R \cdot H_0/R_0 (R/R_0)^{2/7}.
\end{eqnarray}
We fix $H_0$ = 40 AU, leaving us with two free parameters for the disk, the surface density $\Sigma_0$ at $R_0$ and the outer radius where the density profile (Eq.~\ref{eq:disk}) is truncated. The inner radius of the disk has been fixed at 1 AU. This choice may to some extent affect the appearance of the SED at near-infrared wavelengths, but this has no influence on the parameter values we derive in this paper. We did not include the disk inclination as a free parameter, but rather adopted a value for IRS~63 of 30$^\circ$ from \citet{2008A&A...481..141L} who based that number on the brightness of the 3-5 $\mu$m fluxes. Note also that in order to see the ring structure in the visibility amplitudes as described in Sect.~\ref{results}, the disk needs to be relatively face-on. We did however test SEDs calculated for both higher and lower inclinations and found that while any fit is largely indistinguishable at inclinations lower than 30$^\circ$, the fits become rapidly worse at inclinations higher than 50$^\circ$. We did not have a previous estimate for the inclination for IRS~43, but we adopted a high inclination of 70$^\circ$ based on the flattened appearance of the HCO$^+$ moment map (Fig.~\ref{dataplot}) and the fact that we do not see any sign of a ring pattern in the continuum visibility amplitudes.

For the envelope we use a simple spherical power-law model,
\begin{eqnarray}
\rho_{env} (r) = \rho_0 (r/r_0)^{-p},
\end{eqnarray}
with four free parameters, the density $\rho_0$ at $R_0$, the power-law slope $p$ and the inner and outer radius of the envelope. The total density is given by the sum of $\rho_{disk}$ and $\rho_{env}$. 

With these profiles we have eight free parameters in total. The temperature is calculated self-consistently using the radiation transfer code RADMC-3D\footnote{http://www.ita.uni-heidelberg.de/$\sim$dullemond/software/radmc-3d/} and opacities of coagulated dust with thin ice mantles from \citet{Ossenkopf1994}.  We use the LIME radiation transfer code \citep{2010A&A...523A..25B} to calculate continuum images at 450 $\mu$m and 850 $\mu$m, and 1.1 mm. The latter is sampled by the visibilities from our SMA observations and (u,v)-amplitudes are extracted from the resulting visibility set using MIRIAD. The model fluxes from RADMC-3D and the images and visibilities from LIME are compared simultaneously to the SED, the SCUBA images and the SMA 1.1 mm visibility amplitudes. The eight input parameters are varied to obtain the best fitting model. We use the PIKAIA genetic algorithm \citep{Charbonneau1995} to search the parameter space and once the best fit has been found, a local gradient search algorithm is used to estimate the error bars on the parameter values. We assume that the errors in the data are dominated by a 20\% calibration uncertainty. We ran more than hundred thousand SED models for the PIKAIA algorithm to converge on the best solution and we ran the optimization scheme twice to make sure that the same solution was obtained using a different random number seed. The error bars on the parameter values are determined by the distance along the axis in parameter space where the $\chi^2$ value has increased by one with respect to the best fit $\chi^2$ value.

\begin{figure}
   \centering
   \includegraphics[width=8.5cm]{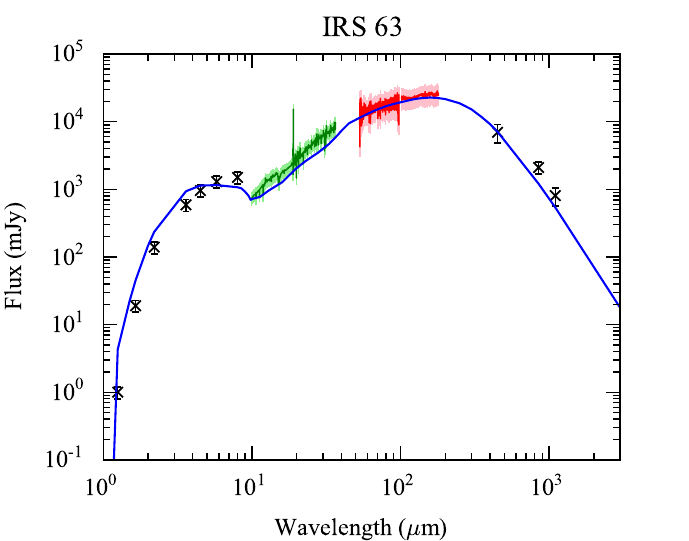}
   \caption{SED of IRS~63 with the best fit model shown with a blue line. Each flux point is shown with a 20\% calibration error bar. The green spectrum is from the Spitzer Space Telescope and the red spectrum is from the Herschel Space Observatory. The near-infrared points are observations from 2MASS and Spitzer/IRAC. The submillimeter data points are from SCUBA (450 $\mu$m and 850 $\mu$m) and CSO (1.1 mm).}
   \label{sed_irs63}
\end{figure}

\begin{figure*}
   \centering
   \includegraphics[width=5cm]{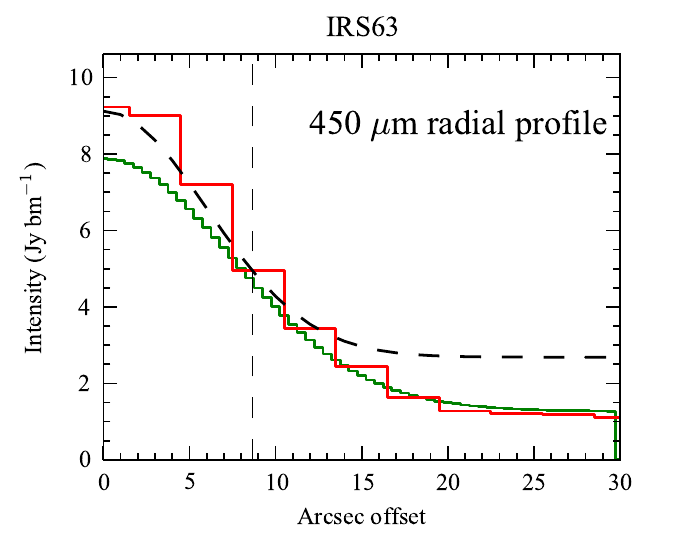} 
   \includegraphics[width=5cm]{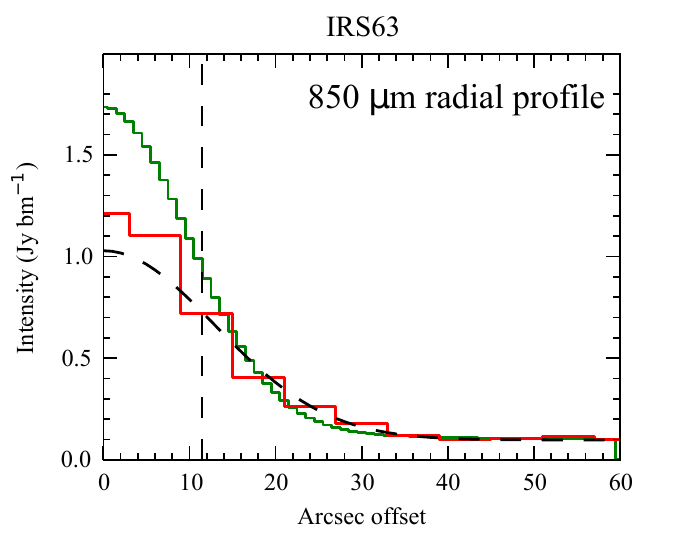}
   \includegraphics[width=5cm]{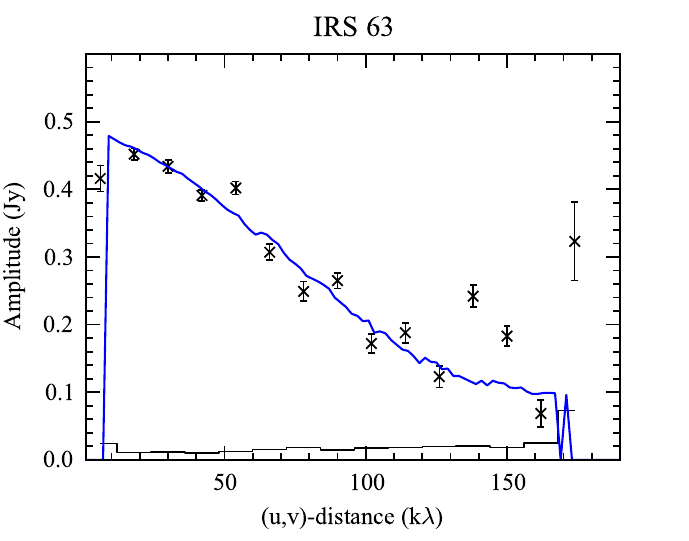} \\
   \includegraphics[width=5cm]{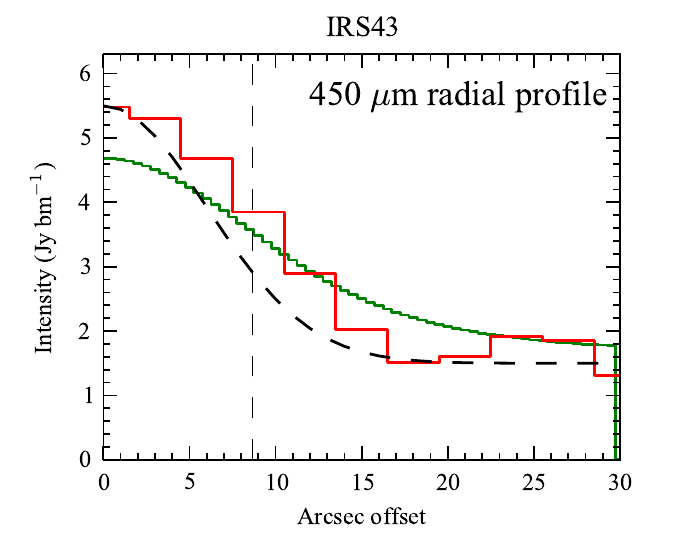} 
   \includegraphics[width=5cm]{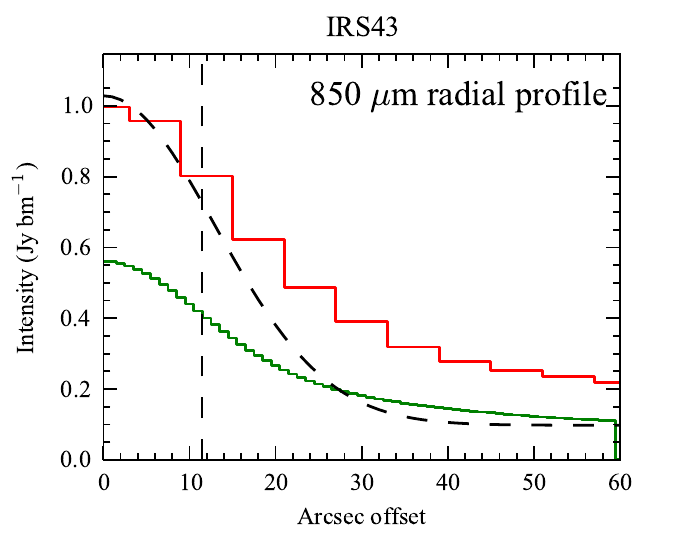}
   \includegraphics[width=5cm]{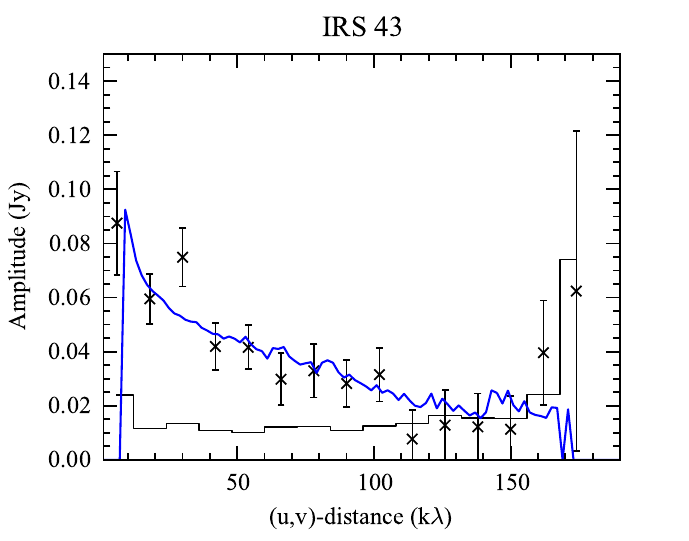}

   \caption{Continuum model of IRS~63 (top row) and IRS~43 (bottom row). The panels show from left to right the model fit to the radial brightness distribution of the SCUBA images and the model fit to the 1.1 mm SMA averaged visibility amplitudes. In the SCUBA profile plots, the data are shown with a red histogram and the model with a green histogram. The black dashed curve is a 2D Gaussian fit to the data. The vertical dashed line indicates the beam size. The visibility amplitudes of IRS~63 are the same as are shown in Fig.~\ref{ring} but binned in much wider bins. In the visibility plots, the Fourier transform of the model is shown in blue.}
   \label{contin_irs63}
\end{figure*}

First we consider IRS~63. Including the Spitzer IRS and Herschel PACS spectra for IRS~63, we have a fully sampled SED from 10 to 200 $\mu$m. The resulting best SED fit is shown in Fig.~\ref{sed_irs63} and the corresponding best fits to the SCUBA and SMA data are shown in Fig.~\ref{contin_irs63}. The comparison between our model and the SMA data is done in (u,v)-space: after multiplication with the primary beam of the SMA and Fourier transformation, the model is sampled with the observed visibilities. Table~\ref{irs63} shows the best fit parameters for IRS~63, including error bars. The disk surface density and envelope reference density are given at the disk radius and the outer radius of the envelope respectively. Interestingly, we find a best fitting disk radius of 165 AU which is almost the same radius as the ring we identified in the continuum visibilities in Sect.~\ref{results}, even though we binned the visibilities in much wider bins to smooth out the nulls during the model optimization.\\

\begin{table}
      \caption[]{Best fit continuum model parameters and 1$\sigma$ error bars (for IRS~63).}
         \label{irs63}
$$
         \begin{array}{p{0.5\linewidth}l l}
            \hline
            \noalign{\smallskip}
            Fitted parameters      &   IRS~63 & IRS~43\\
            \noalign{\smallskip}
            \hline
            \noalign{\smallskip}
            Stellar radius [R$_\odot$] 		     & 24  \pm 5      & 24     \\
            Stellar temperature [K]   	         & 1170  \pm 100  & 1500   \\
            Disk outer radius [AU]     		         & 165 \pm 5      & 190    \\
            Disk surface density$^{a,b}$ [g cm$^{-2}$]& 0.06 \pm 0.01 & 0.0015 \\
            Envelope inner radius [AU]           & 19.5 \pm 15    & 19.5   \\
            Envelope outer radius [AU]            & 8000          & 8000   \\
            Envelope density$^{b,c}$ [cm$^{-3}$]& 2.6 \pm 1.1 \times 10^{6} & 4.5\times 10^6\\
            Envelope power-law slope              & -1.4 \pm 0.25 & -1.4   \\
            \noalign{\smallskip}
            \hline
         \end{array}
$$
	\tiny $^a$ of the dust, $^b$ at outer radius, $^c$ of the gas

\end{table}

\begin{figure}
   \centering
   \includegraphics[width=8.5cm]{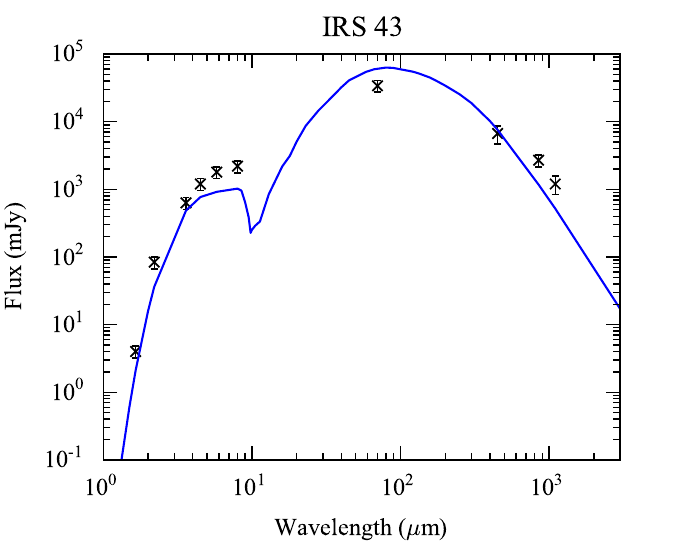}
   \caption{Same as Fig.~\ref{sed_irs63} but for IRS~43.}
   \label{sed_irs43}
\end{figure}

For IRS~43, there is neither a Spitzer nor a Herschel spectrum. The SED of IRS~43 is therefore rather sparse, with only a single MIPS 70 $\mu$m flux point to constrain the mid-infrared part. Correspondingly, PIKAIA could not find a unique solution. The SEDs of IRS~63 and IRS~43, however, are very similar in shape. Therefore, we chose to adopt the same model that fits IRS~63 and simply do a small adjustment to the luminosity and envelope mass to accommodate the difference in absolute flux and then fine tune the resulting solution with a gradient search algorithm to find the local minimum. The best fits to the SED and to the sub-millimeter data can be seen in Fig.~\ref{contin_irs63} and~\ref{sed_irs43}, respectively, and the corresponding parameter values are given in Table~\ref{irs63}. The solution works surprisingly well, suggesting that the two sources are indeed very similar in nature. The main discrepancy between data and the model is at 850 $\mu$m where the peak flux is off by a factor of two. As can be seen in the 850 $\mu$m SCUBA image, however, the emission is contaminated by a neighboring source offset by 35$''$ and therefore the 20\% error bar on the 850 $\mu$m flux point on the SED is probably much too small. Because of the lack of a unique solution for IRS~43, we do not give error bars on the parameter values.

\subsection{The spectral lines}
Once the physical models of IRS~63 and IRS~43 are constrained by continuum emission, we model the HCO$^+$ $J=$3--2 emission lines. At first, we assume a constant abundance for HCO$^+$ which only requires one additional free parameter.  The velocity model that we use is the one first introduced by \citet{brinch2007a}. This model assumes an average velocity field that is spanned by a linear combination of pure free infall and pure Keplerian rotation. The two free parameters are the central (stellar) mass and the ratio of the two basis vectors or rather the angle of the resulting velocity vectors with respect to the azimuthal direction,
\begin{eqnarray}\label{velocity}
\mathbf{v}(\mathbf{r})=\left (
\begin{array}{c}
	v_r \\ 
	v_\phi
\end{array}
\right ) = \sqrt{\frac{GM_*}{r}}
\left (
\begin{array}{c}
	-\sqrt{2} \sin{\alpha} \\ 
	\cos{\alpha}
\end{array}
\right ),
\end{eqnarray}
where $\alpha$ is the angle between $\mathbf{v}(\mathbf{r})$ and the unit vector $\mathbf{\hat{v}}_\phi(\mathbf{r})$. 

\begin{figure*}
   \centering
   \includegraphics[width=8cm]{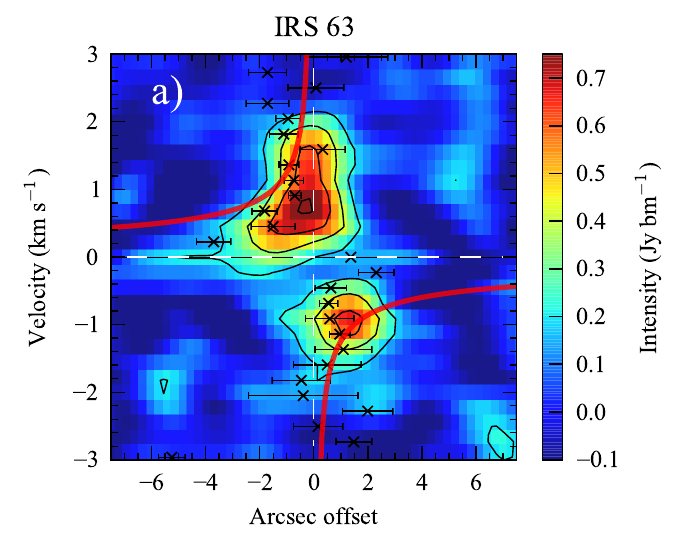} 
   \includegraphics[width=8cm]{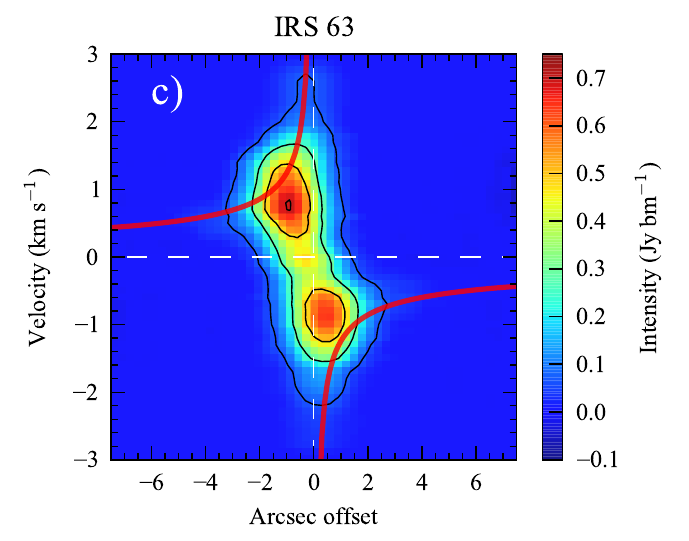}\\   
   \includegraphics[width=8cm]{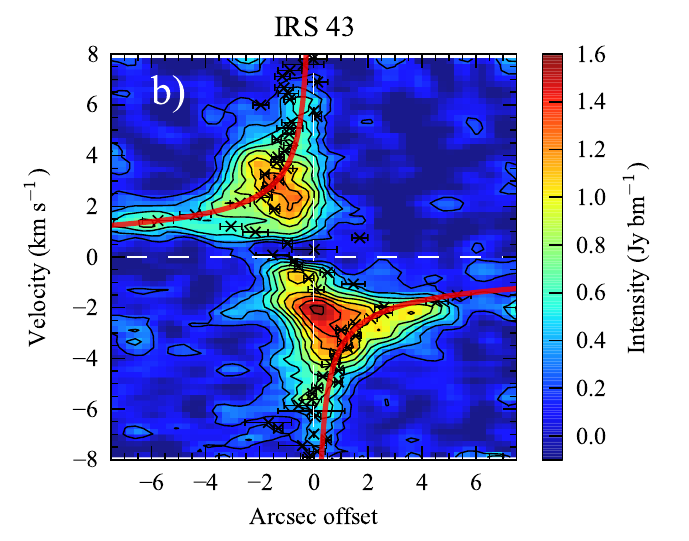} 
   \includegraphics[width=8cm]{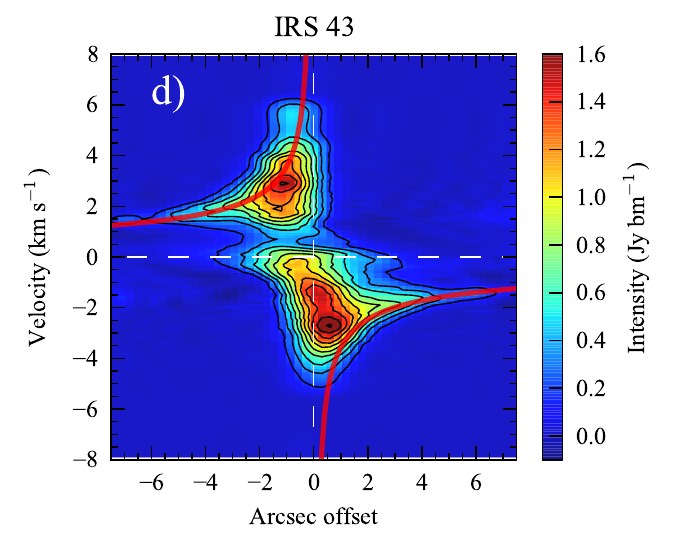} 
   \caption{PV-diagrams of IRS~63 a) and IRS~43 b) and the coresponding best fit models c) and d). The black points and error bars in a) and b) are the channel-by-channel Gaussian fit to the (u,v)-flux. The red curves are not Keplerian rotation, but rather the best fit velocity field as given by Eq.~\ref{velocity}. The PV-diagrams a) and b) are taken along the grey line shown in Fig.~\ref{dataplot}.}
   \label{pv}
\end{figure*}

We pass the physical model including this velocity field to the molecular excitation and radiation transfer code LIME, which creates synthetic spectral image cubes. These are post-processed with MIRIAD to construct synthetic observations that can be compared directly to our SMA data. We used collision rates between HCO$^+$ and H$_2$ from \citet{1999MNRAS.305..651F} taken from the LAMDA database~\citep{Schoier2005}. We also fix the turbulent line broadening at 200 m s$^{-1}$. Again we run the optimization scheme but here constrain the parameters based on the PV-diagrams.  We evaluate our fit in (u,v)-space rather than in the image plane. We thus construct the equivalent to an ordinary PV-diagram directly from the (u,v)-data by fitting a Gaussian to the (u,v)-flux on a channel-by-channel basis. The centroid of these Gaussians form a one-dimensional PV-diagram as can be seen in Fig.~\ref{pv} a) and b) where they are plotted as black crosses on top of the contoured PV-diagram extracted from the image plane.  The distribution of emission is reasonably well reproduced, but in the case of IRS~43, we miss some of the high velocity emission in the wings of the central spectrum at $\pm$6-7 km s$^{-1}$, which could be due to a small contribution of high velocity outflow.  

Panels c) and d) in Fig.~\ref{pv} show our best fit models and the red curves show the best fit radial velocity distribution.  The parameters of the fit are given in Table~\ref{veloparams}. While the velocity model for IRS~43 fits the data almost perfectly, the model for IRS~63 overproduces the velocities between offsets of $\pm 1-3''$. This difference could indicate that we over-estimate either the central mass or the inclination. Lowering either the mass or the inclination makes the lines too narrow and single peaked (the data clearly show a double peak toward the central position), and therefore our best fit in the case of IRS~63 is a trade-off between reproducing the correct line shape and fitting the radial velocity profile. It should be noted that \citet{2008A&A...481..141L} find a stellar mass for IRS~63 which is about half of our value. Using their value makes our model consistent with the emission peak points seen in panel a) of Fig.~\ref{pv}, but the emission distribution is inconsistent. 
\begin{table}
      \caption[]{Best fit line model parameters for IRS~63 and IRS~43.}
         \label{veloparams}
$$
         \begin{array}{p{0.3\linewidth} l l}
            \hline
            \noalign{\smallskip}
            Fitted parameters      & IRS 63  & IRS 43  \\
            \noalign{\smallskip}
            \hline
            \noalign{\smallskip}
            Stellar mass [M$_\odot$] 		  & 0.8    & 1.9   \\
            Inclination    	     			  &  30^\circ      & 70^\circ   \\
            $\alpha \equiv \arctan(v_r/v_\phi)$   &  6^\circ & 16^\circ    \\
	   HCO$^+$/H$_2$ constant           & 0.1 \times 10^{-9} & 0.9 \times 10^{-9} \\
	   HCO$^+$/H$_2$ drop                 & 0.9 \times 10^{-9} & 0.9 \times 10^{-9} \\
            \noalign{\smallskip}
            \hline
         \end{array}
$$
\end{table}

A constant HCO$^+$ abundance with respect to H$_2$ fits the data well for both sources. The value, however, differs by almost a factor of 10 between the two sources. This difference can be explained by freeze-out of the molecules at low temperatures, particularly in the disk mid-plane and in the outer envelope. New models where the molecules are allowed to freeze-out by lowering the abundance by a factor of 10 at temperatures below 30~K produces an equally good fit to the PV-diagrams with an adjustment to the gas phase abundance. We find that while IRS~43 is hardly affected by the freeze-out and thus requires the same gas-phase abundance of 0.9$\times 10^{-9}$ with respect to H$_2$, the gas-phase abundance in IRS~63 needs to be adjusted by a large factor to make the model fit. It turns out that with a freeze out temperature of 30 K, IRS~63 needs to have exactly the same gas phase abundance as IRS~43. The explanation for this difference between the two sources is that the more massive disk in IRS~63 is much colder that the disk in IRS~43, and thus many more molecules are frozen out.   
\begin{table*}
      \caption[]{Comparison of sources}
         \label{derived_pars}
$$
         \begin{array}{p{0.1\linewidth}l c c c c c c }
            \hline
            \noalign{\smallskip}
                  &  Region & L_{bol} [L_\odot] & M_{env} [M_\odot] & M_{disk} [M_\odot] & M_{star} [M_\odot] & Inclination \\
            \noalign{\smallskip}
            \hline
            \noalign{\smallskip}
            IRS 63 & Ophiuchus            & 1.0  & 0.07   & 0.099   & 0.8  & 30^\circ  \\
	    IRS 43 & Ophiuchus            & 2.6  & 0.22   & 0.004   & 1.9  & 70^\circ   \\
	    L1489~IRS$^a$& Taurus & 3.7 & 0.09    & 0.004   & 1.4  & 74^\circ  \\
            \noalign{\smallskip}
            \hline
         \end{array}
$$
	    \tiny{$^a$ Numbers are taken from \citep{2007A&A...475..915B}}.

\end{table*}

Table~\ref{derived_pars} shows properties derived from our best fit models. We also include the corresponding properties of Taurus Class I source L1489~IRS for comparison \citep{2007A&A...475..915B}.

\section{Discussion}
We find that IRS~43 and IRS~63 are physically very similar to each other, in terms of their SEDs and their velocity fields. The peculiar difference in the relative strengths of the lines to the continuum is explained by geometry as well as a difference in the disk mass. The difference in disk mass as well as in luminosity explains why IRS~63 is more affected by freeze-out and if these differences are taken into account, the HCO$^+$ abundances are indeed similar. This is likely a consequence of CO freeze-out on dust grains and consequently a drop in HCO$^+$ as this species follows that of CO closely in protostars \citep{Jorgensen2004c}. 

\begin{figure*}
   \centering
   \includegraphics[width=6.0cm]{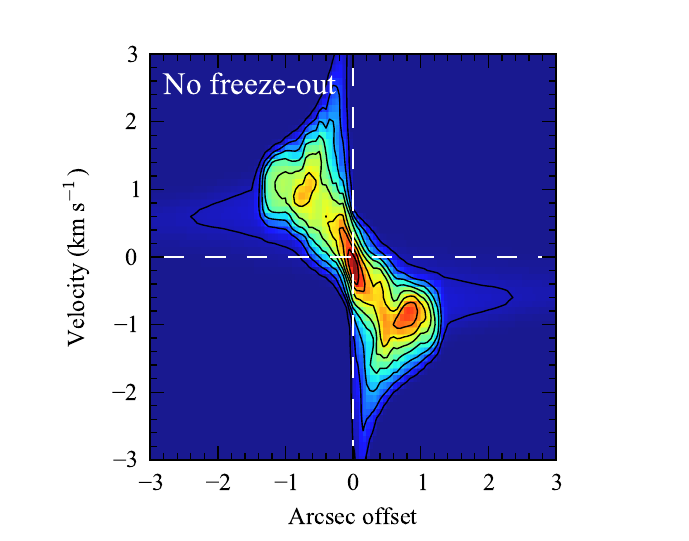}
   \includegraphics[width=6.0cm]{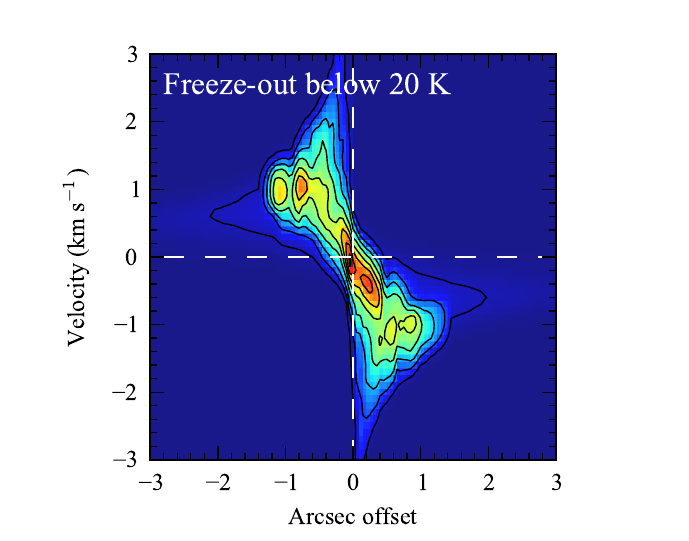}
   \includegraphics[width=6.0cm]{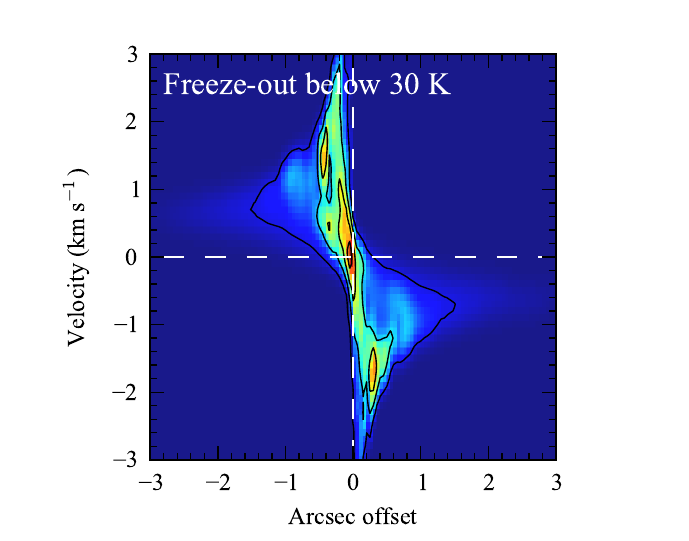}
   \caption{PV-diagrams of the IRS~63 model in a high spatial resolution of 0.05$''$. In the left panel, the HCO$^+$ abundance is constant and in the center and the right panel, the HCO$^+$ abundance is an order of magnitude lower at temperatures below 20 K and 30 K respectively.}
   \label{snow}
\end{figure*}

Various attempts in the recent literature have been made to measure the CO snow line using ALMA~\citep{2013A&A...557A.132M,2013Sci...341..630Q}. Given ALMA resolution, observations of HCO$^+$ in IRS~63 could possibly also reveal the location of the CO snow line in this early disk. Figure~\ref{snow} shows our best fit model of IRS~63 with and without freeze-out below 30 K in a high ALMA resolution of 0.05$''$. While we do not see any distinguishable difference in the PV-diagrams between the freeze-out and no freeze-out cases with the resolution of the SMA (except for the absolute line intensity), there is a clear and distinguishable difference between the two cases when we go to a resolution of 0.05$''$, at least when the freeze-out temperature is 30 K. For a freeze-out temperature of 20 K, the depleted region is too small to have any impact on the overall emission. 

We find that our best fit model parameters are close to the values derived in \citet{Jorgensen:2009bx}. Our model results on IRS~63 are also mostly consistent with the parameters derived by \citet{2008A&A...481..141L} except for the stellar mass. They, however, base their stellar mass on the PV-diagram alone (and with much sparser data) which, as discussed above, may lead to an inaccurate stellar mass. We thus conclude that envelope and disk masses and stellar luminosities of Class I sources can be safely derived directly from observations by measuring the total and the compact submillimeter flux and integrating the SED, as done by \citet{Jorgensen:2009bx}.  

\begin{figure}
   \centering
      \includegraphics[width=8.5cm]{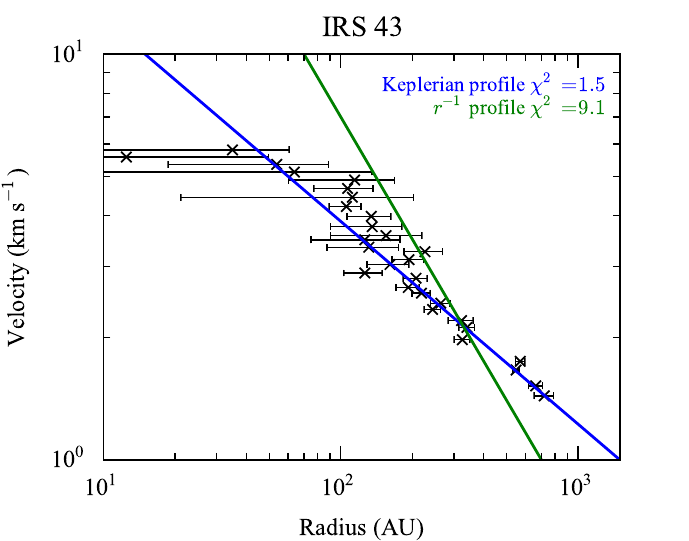}
   \caption{The emission peaks per velocity channel from the PV-diagram of IRS~43 in panel b in Fig.~\ref{pv}, plotted in logarithmic space. The blue line shows our best fit velocity model and the green line shows the best $r^{-1}$ profile that fits the data.}
\label{veloprof}
\end{figure}

Currently, there is some discussion in the literature on whether Class I disks are truly Keplerian or whether the velocity field in these disks are consistent with other profiles too~\citep{2013arXiv1305.0627B,2013ApJ...772...22Y}. As illustrated in Fig.~\ref{veloprof}, the velocity field of IRS~43, is in almost perfect agreement with a Keplerian $r^{-0.5}$ profile . For comparison, we have also plotted the best possible $r^{-1}$ profile which does not agree at all with the data. If we consider the IRS~63 velocity data in the same way, we get a much less clear picture. It is in fact possible to fit both profiles with about the same $\chi^2$. This indeterminacy, however, does not mean that IRS~63 does not have a Keplerian disk but is simply a reflection of the fact that an almost face-on disk shows very weak rotation signature and that the signal-to-noise ratios of the IRS~63 line are not high. From a purely geometrical argument, about 15\% of all disks have an inclination of 30$^\circ$ (like IRS~63) or lower, which means that \emph{at least} 15\% of all Class I objects with Keplerian disks may not show a strong, unique Keplerian velocity profile. 

Figure~\ref{snow} shows that freeze-out has the effect of removing emission from larger angular offsets. Figure~\ref{veloprof} shows that it is exactly that emission which constrains the velocity profile and makes us able to distinguish Keplerian rotation from a $r^{-1}$ velocity profile. It is therefore possible to mistake a $r^{-0.5}$ profile for an $r^{-1}$ profile in a Class I object, if one is using a tracer which is depleted, no matter the spatial resolution. 

While the two sources appear rather similar, it is clear that IRS~63 has a more massive disk and less envelope left than IRS~43. This difference could be interpreted as one being more evolved that than the other. Both sources, however, only have about 10\% mass left in the envelope and both are dominated by Keplerian rotation which means that they must both be close to the T Tauri stage. Although some very high velocity material is seen in the spectrum of IRS~43, neither source shows any substantial outflow.

\section{Summary}
In this paper we present new high angular resolution observations of HCO$^+$ J=3-2 and the continuum at 1.1 mm of two Class I sources in Ophiuchus, IRS~63 and IRS~43. We perform detailed radiative transfer modeling of the dust in order to reproduce the SED as well as continuum images at 450 $\mu$m, 850 $\mu$m, and 1.1 mm. We go on to model the HCO$^+$ line using a non-LTE radiative transfer method. We draw a number of conclusions based on this modeling:
\begin{itemize}
\item We have identified a ring structure in the 1.1 mm continuum visibilities of IRS~63. No such structure is seen in IRS~43 and we conclude that this difference is due to the difference in inclination.  The ring coincides with the modeled disk radius, but our model fails to reproduce the signature in the (u,v)-plane. The ring does not appear in the image plane and the nature and origin of it is still an open question. Higher resolution and sensitivity observations with ALMA can potentially reveal the nature of this curious structure around the edge of the disk.\\ 

\item We show that the velocity field of IRS~43 is very well described by a Keplerian velocity field on scales between ~10 and 700 AU. The case for Keplerian motion is not equally clear for IRS~63, due to the lack of signal-to-noise, but this is consistent with the fact that IRS~63 is seen more or less face-on whereas IRS~43 is much closer to edge-on.\\

\item We find that our best-fit model parameters are consistent with the results of previous studies, except for the case of the central mass of IRS~63, which we find was previously underestimated. We furthermore find that the freeze-out chemistry is important for IRS~63 and relatively unimportant for IRS~43 -- likely a consequence of the more massive disk and less luminous central source of IRS~63 as compared to IRS~43. Table~\ref{derived_pars} shows that IRS~43 and IRS~63 are very similar to L1489~IRS, which has previously been claimed to be a unique source~\citep{Hogerheijde2001}. In particular, IRS~43 seems almost identical, especially when comparing their morphologies.

\end{itemize}

\noindent \emph{Acknowledgments:} This research was supported by a grant from the Carlsberg Foundation to Christian Brinch and by a grant from the Lundbeck Foundation Group Leader Fellowship and by a grant from the Instrumentcenter for Danish Astrophysics (IDA) to Jes J{\o}rgensen. Research at Centre for Star and Planet Formation is funded by the Danish National Research Foundation and the University of Copenhagen's programme of excellence.
\\

\bibliographystyle{aa}
\bibliography{references,bergin}

\end{document}